\documentclass{webofc}
\usepackage[varg]{txfonts}
\usepackage[utf8]{inputenc} 
\usepackage{graphicx}
\usepackage{dcolumn}
\usepackage{bm}
\usepackage[bookmarks, pagebackref=false]{hyperref}
\usepackage[nameinlink]{cleveref}
\usepackage[usenames,dvipsnames]{xcolor}
\usepackage[normalem]{ulem}
\usepackage{physics}
\definecolor{rossoCP3}{cmyk}{0,.88,.77,.40}
\definecolor{blaa}{rgb}{0.2,0.2,0.6}
\hypersetup{colorlinks, 
	bookmarksopen, 
	bookmarksnumbered,
	citecolor=blaa, 		
	linkcolor=rossoCP3,	
	urlcolor=rossoCP3,			
}

\crefname{section}{Sec.\!}{Secs.\!}
\crefname{figure}{Fig.\!}{Figs.\!}
\crefname{equation}{}{}
\crefname{table}{Tab.\!}{Tabs.\!}
\crefname{appendix}{App.\!}{Apps.\!}

\usepackage{natbib}
\usepackage{slashed}
\newcolumntype{x}[1]{>{\centering\arraybackslash\hspace{0pt}}p{#1}}

\graphicspath{{./figs/}}

\begin{document}
 
\title{Gravitational Waves from dark composite dynamics}

\author{\firstname{Manuel} \lastname{Reichert}\inst{1}\fnsep\thanks{\email{m.reichert@sussex.ac.uk}}
	 \and
\firstname{Zhi-Wei} \lastname{Wang}\inst{2,3}\fnsep\thanks{\email{zhiwei.wang@thep.lu.se}}}

\institute{Department  of  Physics  and  Astronomy,  University  of  Sussex,  Brighton,  BN1  9QH,  U.K
	\and
	School of Physics, The University of Electronic Science and Technology of China,\\
 88 Tian-run Road, Chengdu, China
\and
Department of Astronomy and Theoretical Physics, Lund University
}

\abstract{
We discuss the stochastic gravitational-wave spectrum from dark confinement and chiral phase transitions in the early Universe. Specifically, we look at pure Yang-Mills theory for an arbitrary number of colours as well as SU(3) with quarks in different representations. We utilise thermodynamic Lattice data and map it to effective models, such as the Polyakov-loop and the PNJL model. This allows us to compute gravitational-wave parameters and the corresponding gravitational-wave signal. We compare the signal to future gravitational-wave observatories such as the Big Bang Observer and DECIGO.
} 
 
\maketitle

\section{Introduction}
\nocite{Huang:2020crf} 
\nocite{Reichert:2021cvs} 

Very little is known about the dark side of the Universe and it is therefore highly desirable to be able to test the immense landscape of dark sectors. Here, we discuss the scenario that the dark side features composite sectors made by non-Abelian Yang-Mills theories. These theories are physically motivated because the dynamics of the dark sector naturally mimics the Standard-Model (SM) QCD featuring strong interactions. Furthermore, these theories are well-behaved at short distances due to asymptotic freedom, meaning that the theories are, per se, ultraviolet complete and they do not introduce new types of hierarchies beyond the SM one. These strongly coupled dark sectors have recently received a lot of attention due to their non-trivial mechanisms to create dark-matter bound states \cite{Boddy:2014yra, Hochberg:2014dra, Cacciapaglia:2020kgq, Dondi:2019olm, Asadi:2021yml}. Depending on the portal to the SM, these types of theories are unfortunately inaccessible to current colliders or direct searches, limiting our ability to test them and therefore pin down the model underlying the dark sector. Here we discuss the possibility to test these hidden sectors via the detection of gravitational waves (GWs). Importantly, the GW signal does not depend on the precise portal coupling of the dark sector to the SM and we only need to assume the minimal interaction of gravity between the SM and new strongly-coupled sectors. 

In this contribution, we review the results from \cite{Huang:2020crf,Reichert:2021cvs} where the GW signals from a dark first-order confinement and chiral phase transition were studied. These results include the pure glue SU($N_c$) case as well as the SU(3) case with quarks in the fundamental, adjoint, and two-index symmetric representation. The strongly coupled nature of the dark sectors requires a non-perturbative approach. We use effective models such as the Polyakov-Loop model (PLM) and the Polyakov-Nambu-Jona-Lasinio (PNJL) model fitted to non-perturbative lattice data to get an accurate description of the dynamics of the phase transition. The resulting GW signals are compared to the sensitivity curves of future GW detectors, most prominently LISA, the Big Bang Observer (BBO), and DECIGO. Further approaches to study the GW signal of strongly-coupled sectors include the Matrix Model \cite{Halverson:2020xpg}, and holography \cite{Ares:2021nap, Ares:2021ntv, Morgante:2022zvc, He:2022amv} and the Linear-Sigma model \cite{Helmboldt:2019pan}.

\section{Effective Models}
\label{sec:eff-models}
\subsection{Polyakov Loop}
In any SU($N_c$) gauge theory, a global $Z_{N_c}$ symmetry, called the centre symmetry, naturally emerges from the associated local gauge symmetry. It is possible to construct several gauge invariant operators charged under this global $Z_{N_c}$ symmetry. Among them, the most notable one is the Polyakov loop,
\begin{align}
	{\ell}\left(x\right)=\frac{1}{N_c}{\rm Tr}\left[{\mathcal P}\exp(i\,g\int_{0}^{1/T} \!\! A_{0}(x,\tau)\,\mathrm d\tau)\right]\,,
	\label{eq:Polyakov_Loop}
\end{align}
where  $\cal P$ denotes the path ordering, $g$ is the SU($N_c$) gauge coupling, and $A_0$ is the vector potential in the time direction. The symbols $x$ and $\tau$ denote the three spatial dimensions and the Euclidean time, respectively. An important feature of the Polyakov loop is that its expectation value vanishes below the critical temperature $T_c$, i.e.\ $\langle\ell\rangle_{T<T_c}=0$, while it possesses a finite expectation value above the critical temperature, i.e.\ $\langle\ell\rangle_{T>T_c}>0$ . At very high temperatures, the vacua exhibit a $N_c$-fold degeneracy and we have
\begin{align}
	\langle \ell\rangle&=\exp(i\frac{2\pi j}{N_c})\ell_0\,,
	&
	j&=0,1,\ldots,(N_c-1)\,,
\end{align}
where $\ell_0$ is defined to be real and $\ell_0\rightarrow 1$ as $T\rightarrow\infty$. In summary, the Polyakov loop is a suitable order parameter in the finite temperature phase transition of the SU($N_c$) gauge theory.

\subsection{Effective Potential of the Polyakov-Loop Model}
We use the PLM \cite{Pisarski:2000eq, Pisarski:2001pe} as an effective theory to describe the confinement phase transition. The expectation value of the Polyakov loop \eqref{eq:Polyakov_Loop} plays the role of an order parameter. The simplest effective potential preserving the $Z_{N_c}$ symmetry is given by
\begin{align} \label{eq:PLM_full}	
	V_{\rm{PLM}} &=T^4\left(-\frac{b_2(T)}{2}|\ell|^2+b_4|\ell|^4-b_3\!\left(\ell^{N_c}+\ell^{*N_c}\right)\right),\notag \\
	b_2(T)&=a_0+a_1\!\left(\frac{T_c}{T}\right)\!+a_2\!\left(\frac{T_c}{T}\right)^{\!2}\!+a_3\!\left(\frac{T_c}{T}\right)^{\!3}\!+a_4\!\left(\frac{T_c}{T}\right)^{\!4}.
\end{align}
We have chosen the coefficients $b_3$ and $b_4$  to be temperature independent following the treatment in \cite{Ratti:2005jh, Fukushima:2017csk}, which studied the SU(3) case, and also neglected higher orders in $\lvert\ell\rvert$.

The coefficients in \eqref{eq:PLM_full}  are fitted to the thermodynamic lattice data from \cite{Panero:2009tv}. In particular, we are using the energy and entropy density and employ a $\chi^2$ analysis. We fit the data for $N_c=3,4,5,6,8$. For example for $N_c =3$, we find the best fit values $a_0=3.72$, $a_1= -5.73$, $a_2= 8.49$, $a_3=-9.29$, $a_4 = 0.27$, $b_3 = 2.4$, and $b_4 = 4.53$. The values for other $N_c$ are reported in Tab.~1 in \cite{Huang:2020crf}.

\subsection{Polyakov-Nambu-Jona-Lasinio Model}
\label{sec:PNJL}
The PNJL model is used to describe dynamics in dark gauge-fermion sectors \cite{Fukushima:2017csk}.
The finite-temperature grand potential of the PNJL models can be generically written as 
\begin{align}
	V_{\rm{PNJL}}=V_{\rm{PLM}}[\ell,\ell^*]+V_{\rm{cond}}\!\left[\langle \bar{\psi}\psi\rangle\right]+V_{\rm{zero}}\!\left[\langle \bar{\psi}\psi\rangle\right]
	+V_{\rm{medium}}\!\left[\langle \bar{\psi}\psi\rangle, \ell,\ell^*\right]\,,
\end{align}
where $V_{\rm{PLM}}$, $V_{\rm{cond}}$, and $V_{\rm{zero}}$ denotes respectively the Polyakov-loop potential (discussed above), the condensate energy and the fermion zero-point energy. The medium potential $V_{\rm{medium}}$ encodes the interactions between the chiral and gauge sector which arises from an integration over the quark fields coupled to a background gauge field. We employ the PNJL model for SU(3) gauge theories with $N_f=3$ fundamental quarks, $N_f=1$ adjoint quarks, and $N_f=1$ two-index symmetric quarks.

The explicit form of these contributions is detailed in \cite{Reichert:2021cvs}. They contain new free parameters such as the coupling of the four-fermion interaction $G_S$, the six-fermion interaction $G_D$ from the Kobayashi-Maskawa-’t Hooft determinantal term, and the cutoff scale $\Lambda$. These parameters can be fixed by matching observables such as the constituent quark mass, the pion-decay constant, and the $\sigma$-meson mass. We fix these parameters in two ways: We have a benchmark point where we use the values from \cite{Fukushima:2017csk, Kahara:2012yr} as guidance, rescaled to a confinement temperature $T_c = 100$\,GeV. We furthermore perform a parameter scan and determine the strongest possible phase transition. The values for all cases and parameters are displayed in Tab.~3 in \cite{Reichert:2021cvs}.

\section{First-order Phase Transitions and Gravitational Waves}
\label{sec:PhaseTransition}
We compute GW parameters such as the percolation temperature $T_p$, the energy budget $\alpha$, and the inverse duration time $\beta$, which describe the dynamics of the bubble-nucleation process. The wall velocity $v_w$ is difficult to access and therefore we leave it as a free input parameter.

\subsection{Bubble Nucleation}
\label{sec:bubble-nucleation}
In the conventional picture of a first-order phase transition, the universe cools down and a second minimum with a non-zero vacuum expectation value (broken phase) develops at a critical temperature. This triggers the tunnelling from the false vacuum (unbroken phase) to the stable vacuum (broken phase) below the critical temperature. In the case of the confinement phase transition, the picture is reversed: as the universe cools down, the tunnelling occurs from the broken phase (deconfinement phase) to the unbroken phase (confinement phase), since the underlying discrete symmetry $Z_{N_c}$ is broken in the deconfinement phase at high temperature while it is preserved at the confinement phase at low temperature (so-called symmetry non-restoration). 

The tunnelling rate  per unit volume due to thermal fluctuations is suppressed by the three-dimensional Euclidean action $S_3(T)$ \cite{Linde:1980tt, Linde:1981zj},
\begin{align}
	\Gamma(T)=T^4\left(\frac{S_3(T)}{2\pi T}\right)^{\!3/2} e^{-S_3(T)/T}.
	\label{eq:decay_rate}
\end{align}
The three-dimensional Euclidean action for a confinement transition in the Polyakov loop reads
\begin{align}
	S_3(T)=4\pi T \!\int_0^\infty \!\!\mathrm dr'\, r'^2\!&\left[\frac{1}{2}\left(\frac{\mathrm d\ell}{\mathrm dr'}\right)^{\!2} +V'_\text{eff}(\ell,T)\right]\,,
	\label{eq:Euclidean_Action}
\end{align}
where $V'_\text{eff} (\ell,T) \equiv V_\text{eff}(\ell,T)/T^4$ and $r'=r\, T$ are dimensionless. The bubble profile (instanton solution) is obtained by solving the equation of motion of the action in \eqref{eq:Euclidean_Action}
\begin{align}
	\frac{\mathrm d^2\ell(r')}{\mathrm dr'^2}+\frac{2}{r'}\frac{\mathrm d\ell(r')}{\mathrm dr'}-\frac{\partial V_\text{eff}'(\ell,T)}{\partial\ell}=0\,,
	\label{eq:bounce-solution}
\end{align}
with the associated boundary conditions
\begin{align}
	\frac{\mathrm d\ell(r'=0,T)}{\mathrm dr'}&=0\,,
	&
	\lim_{r'\rightarrow 0} \ell(r',T)&=0\,.
	\label{boundary_confinement}
\end{align}
To attain the solutions, we used the method of overshooting/undershooting and employ the \texttt{Python} package \texttt{CosmoTransitions} \cite{Wainwright:2011kj}.

For the chiral phase transition, we use the chiral condensate $\sigma \propto \langle \bar \psi \psi \rangle$ as the order parameter for the phase transition. For our models, this applies only to the case with fundamental quarks. In this case, the tunnelling process is conventional i.e.~from the unbroken phase to the broken phase below the critical temperature. The procedure to obtain the bubble profile is in straight analogy to the Polyakov-loop case. However, the mass dimensions in \cref{eq:Euclidean_Action} and the boundary conditions \cref{boundary_confinement} adjusted. Furthermore, we include the wave-function renormalisation $Z_\sigma$ since $\sigma$ is not a fundamental field \cite{Helmboldt:2019pan}, which slightly modifies the equation of motion \cref{eq:bounce-solution}. For more details see \cite{Reichert:2021cvs}.

\subsection{Inverse Duration Time of the Phase Transition}
\label{sec:beta}
An important parameter for the computation of the GW signal is the inverse duration time $\beta$. For sufficiently fast phase transitions, the decay rate can be approximated by $\Gamma(T) \approx \Gamma(t_*) e^{\beta (t-t_*)}$ where $t_*$ is the characteristic time scale for the production of GWs. The dimensionless inverse duration time then follows as
\begin{align}
	\tilde \beta = \frac{\beta}{H_*}=T\frac{\mathrm d}{\mathrm dT}\frac{S_3(T)}{T}\bigg\vert_{T=T_*}\,,
	\label{eq:beta-tilde}
\end{align}
where we used that $\mathrm dT/\mathrm dt = -H(T)T$ with the Hubble parameter $H$. The phase transition temperature $T_*$ is often taken as the nucleation temperature $T_n$, which is defined as the temperature at which the rate of bubble nucleation per Hubble volume and time is approximately one, i.e.\ $\Gamma/H^4\sim \mathcal{O}(1)$. A more accurate definition is to use the percolation temperature $T_p$, which is defined as the temperature at which the probability to have the false vacuum is about $0.7$. For very fast phase transitions, as in our case, the nucleation and percolation temperature are almost identical $T_p\lesssim T_n$. Nonetheless, we use the percolation temperature. We write the false-vacuum probability as $P(T) = e^{-I(T)}$ with the weight function \cite{Ellis:2018mja}
\begin{align}
	I(T)=\frac{4\pi}{3} \int^{T_c}_T \! \!\mathrm dT'\frac{\Gamma(T')}{H(T')T'{}^4} \left( \int^{T'}_{T}\!\!\mathrm dT''\frac{v_w(T'')}{H(T'')}\right)^{\!3}.
	\label{eq:Tp}
\end{align}
The percolation temperature is defined by $I(T_p)=0.34$, corresponding to $P(T_p)= 0.7$. Using $T_*=T_p$ in \eqref{eq:beta-tilde} yields the dimensionless inverse duration time.

Throughout all models we find rather large values of the inverse duration time, $\tilde \beta = \mathcal{O}(10^4)$. In the pure glue case, we find $\tilde \beta  = 1.0 \cdot 10^5$ ($N_c=3$), $\tilde \beta  = 1.3 \cdot 10^5$ ($N_c=4$),  $\tilde \beta  = 1.1 \cdot 10^5$ ($N_c=5$), $\tilde \beta  = 4.2 \cdot 10^4$ ($N_c=6$), and $\tilde \beta  = 4.9 \cdot 10^4$ ($N_c=8$). The smallest value is taken for $N_c=6$ and we therefore expect the strongest GW signal in that case\footnote{The efficiency factor to generate the GW from sound-wave contribution is proportional to $\sqrt{\tau_{\rm{SW}}}$ where $\tau_{\rm{SW}}$ is the sound wave period and $\tau_{\rm{SW}}\sim\frac{1}{\beta}$ for $\beta>>1$. Thus, the larger $\beta$ the more suppression of the GW signals.}. With quarks for SU(3) we find the following results: $\tilde \beta = 1.9\cdot 10^4$ (fundamental, benchmark), $\tilde \beta = 6.8\cdot 10^3$ (fundamental, best case), $\tilde \beta = 8.1\cdot 10^4$ (adjoint, benchmark), $\tilde \beta = 7.9\cdot 10^4$ (adjoint, best case), $\tilde \beta = 5.9\cdot 10^4$ (two-index symmetric, benchmark), and $\tilde \beta = 1.7\cdot 10^4$ (two-index symmetric, best case). We observe that the GW signal from the adjoint phase transition is the weakest.

\subsection{Energy Budget}
We define the strength parameter $\alpha$ from the trace of the energy-momentum tensor $\theta$ weighted by the enthalpy $w$,
\begin{align}
	\alpha=\frac{1}{3}\frac{\Delta\theta}{w_+}=\frac{1}{3}\frac{\Delta e\,-\,3\Delta p}{w_+}\,,
	\label{alpha_def}
\end{align}
where $\Delta X= X^{(+)}-X^{(-)}$ for $X = (\theta$, $e$, $p$) and $(+)$ denotes the meta-stable phase (outside of the bubble) while $(-)$ denotes the stable phase (inside of the bubble). The relations between enthalpy $w$, pressure $p$, and energy $e$ are given by $w=T\partial_T p$ and $e=w - p$. These are hydrodynamic quantities and we work in the approximation where do not solve the hydrodynamic equations but instead extract them from the effective potential $p^{(\pm)}=-V_{\text{eff}}^{(\pm)}$. This treatment should work well for the phase transitions considered here \cite{Giese:2020rtr, Giese:2020znk}. Then $\alpha$ is given by
\begin{align}
	\alpha=\frac{1}{3}\frac{4\Delta V_\text{eff}-T\frac{\partial \Delta V_\text{eff}}{\partial T}}{-T\frac{\partial V_{\text{eff}}^{(+)}}{\partial T}}\,.
	\label{eq:alpha}
\end{align}
In the case of the confinement phase transition, we find that the contribution from $\Delta V_\text{eff}$ is negligible since $e_+\gg p_+$ and therefore $\alpha \approx 1/3$ with at most a 10\% deviation. In the case of the chiral phase transition, i.e., for SU(3) with $N_f=3$ fundamental quarks, we find smaller values, $\alpha = 2.9\cdot 10^{-2}$ (benchmark) and $\alpha = 5.1\cdot 10^{-2}$ (best case). This relates to the fact that there are more relativistic degrees of freedom participating in the phase transition. 

\begin{figure}[t]
	\includegraphics[width=.48\linewidth]{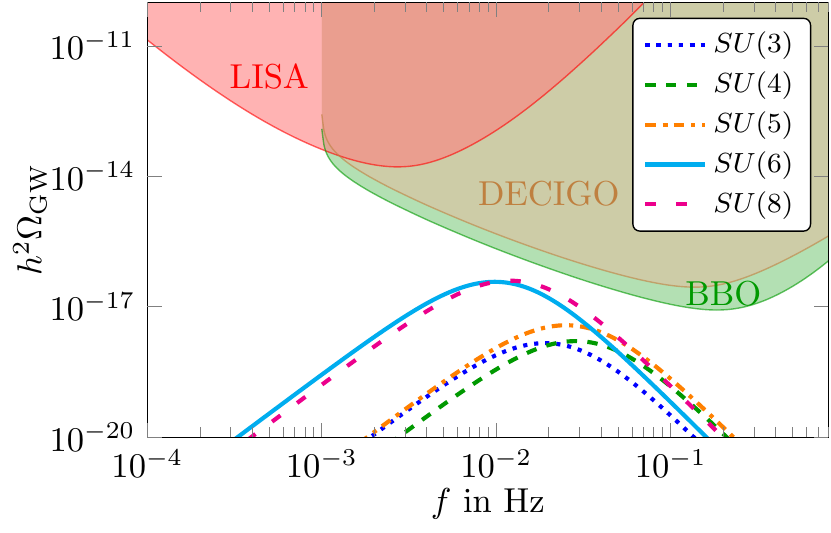}
	\hfill
	\includegraphics[width=.48\linewidth]{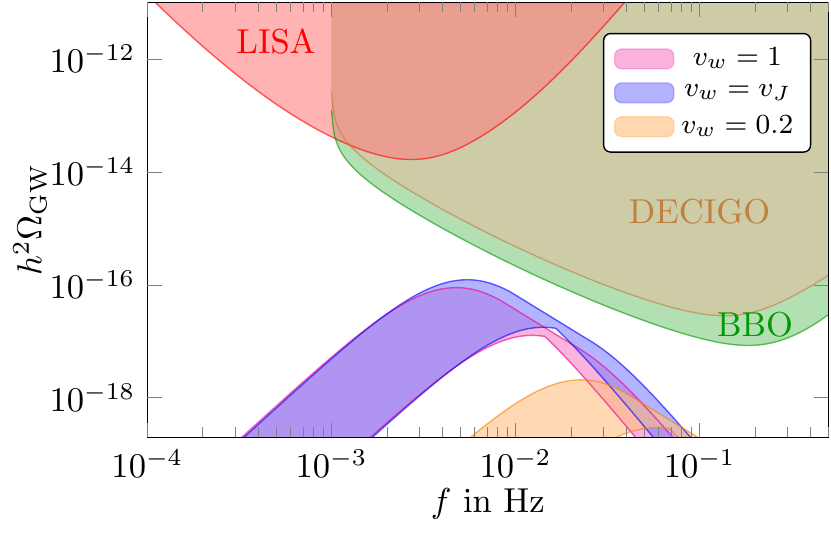}
	\caption{GW spectra from the SU($N_c$) confinement phase transition without quarks for different values of $N_c$ and $v_w= v_J$ (left) as well as for $N_c=6$ and different wall velocities (right). Figures taken from~\cite{Huang:2020crf}.}
	\label{fig:GW1}
\end{figure}

\subsection{Gravitational-Wave Spectrum from Sound Waves}
\label{sec:spectrum}
With the GW parameters, we can compute the GW spectrum using fit formulas from numerical simulations \cite{Caprini:2015zlo, Caprini:2019egz}. In general, there are three contributions to the GW spectrum: collisions of bubble walls, sound waves in the plasma after bubble collision and magnetohydrodynamic turbulence in the plasma. In our case, the contributions from sound waves are dominating. The spectrum, peak frequency, and peak amplitude are given by \cite{Caprini:2015zlo, Caprini:2019egz}
\begin{align} \label{eq:GWsignal} 
	h^2\Omega_\text{GW}(f)&= h^2\Omega^\text{peak}_\text{GW} \left(\frac{f}{f_\text{peak}}\right)^{\!3} \left[ \frac{4}{7}+\frac{3}{7}\left( \frac{f}{f_\text{peak}} \right)^{\!2}\right]^{-\frac{7}{2}},  \notag \\
	f_\text{peak}&\simeq 1.9\cdot 10^{-5}\,\text{Hz}\left(\frac{g_*}{100} \right)^{\!\frac{1}{6}}\left( \frac{T}{100\, \text{GeV}}\right) \left(\frac{\tilde \beta}{v_w} \right), \notag  \\
	h^2\Omega^\text{peak}_\text{GW} &\simeq 2.65\cdot 10^{-6}\left(\frac{v_w}{\tilde \beta}\right)\left( \frac{\kappa\, \alpha}{1+\alpha} \right)^{\!2}\left(\frac{100}{g_*}\right)^{\!\frac{1}{3}}\Omega_\text{SU(N)}^2\,. 
\end{align}
Here, $h= H/(100 \text{km}/\text{s}/\text{Mpc})$ is the dimensionless Hubble parameter and $g_*$ is the effective number of relativistic degrees of freedom including the SM degrees of freedom $g_{*,\text{SM}}=106.75$ and the dark sector ones, e.g., $g_{*,\text{SU($N_c$)}}=2(N_c^2-1)$ in the pure gluon case. The factor $\Omega_\text{SU($N_c$)}=\frac{\rho_\text{rad,SU($N_c$)}}{\rho_\text{rad,tot}}$ accounts for the dilution of the GWs by the visible matter which does not participate in the phase transition.  The efficiency factor $\kappa$ in \eqref{eq:GWsignal} describes which fraction of the energy budget is converted into GWs. We follow the treatment in \cite{Espinosa:2010hh} and also include the additional suppression due to the length of the sound-wave period \cite{Ellis:2019oqb, Ellis:2020awk, Guo:2020grp}.

\section{Gravitational-Wave Spectra}
\label{sec:GravityWaves}
In the previous sections, we presented all the necessary ingredients to compute the GW signal. We first discuss the pure gluon case and then include quarks.

In the left panel of \cref{fig:GW1}, we display the GW spectrum for different $N_c$ and with $T_c = 1$\,GeV and $v_w = v_J$ where $v_J$ is the Chapman-Jouguet detonation velocity. We compare the GW spectra to the power-law integrated sensitivity curves of LISA, BBO, and DECIGO. For a better visibility, we do not include the error on the GW signal in this plot. We observe the largest GW signal for $N_c =6,8$ and a slightly smaller GW signal for $N_c = 3, 4, 5$. The reason for this difference is two-fold: firstly, the GW signals for small $N_c$ are more suppressed by the dilution factor $\Omega_\text{SU($N_c$)}$ in \eqref{eq:GWsignal} since the SM degrees of freedom dominate over the dark degrees of freedom. Secondly, we observe the smallest value of $\tilde \beta$ for $N_c =6$, see \cref{sec:beta}, which makes it the strongest phase transition in the pure glue case.

\begin{figure}[t]
	\includegraphics[width=0.9\linewidth]{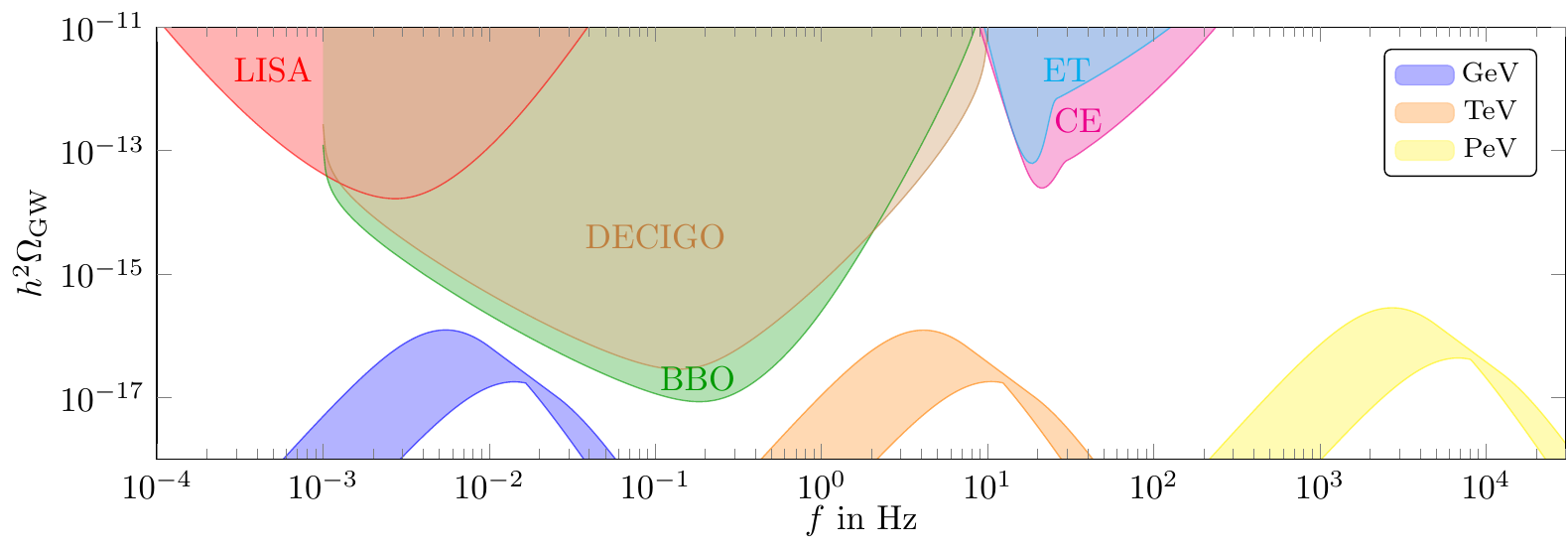}
	\caption{GW spectrum from the SU(6) confinement phase transition without quarks for the confinement scales $T_c = 1$\,GeV, 1\,TeV, and 1\,PeV, compared to the power-law integrated sensitivity curves of different GW detectors. Figure taken from \cite{Huang:2020crf}.}
	\label{fig:GW2}
\end{figure}

In the right panel of \cref{fig:GW1}, we show the GW spectrum for $N_c=6$ and $T_c = 1$\,GeV but vary the wall velocity $v_w$, which we treat as an input parameter. As expected, we observe the largest GW signal for the Chapman-Jouguet detonation velocity. For wall velocities smaller than the speed of sound $v_w \leq c_s = 1/\sqrt{3}$, the GWs become suppressed due to the rapidly decreasing efficiency factor $\kappa$ in \eqref{eq:GWsignal}, see \cite{Cutting:2019zws}. This is exemplified in  \cref{fig:GW1} with $v_w = 0.2$. Furthermore, we also include the errors of the GW signal, which are directly inferred from the fitted lattice data but enhanced with a generous factor of 10 to encompass potential systematic errors from the fitting procedure and the choice of the effective model.

In \cref{fig:GW2}, we display the dependence of the GW spectrum on the confinement temperature $T_c$. We use $N_c=6$ and $v_w = v_J$. The confinement temperature shifts the peak frequency and is thereby one of the key parameters when it comes to the detectability of the GW signal. While the sensitivities of LISA, ET, and CE are not sufficient to test the GW signal, BBO and DECIGO will test dark sectors from roughly $T_c = 5$\,GeV up to $T_c = 100$\,GeV.

\begin{figure}[t]
	\includegraphics[width=.329\linewidth]{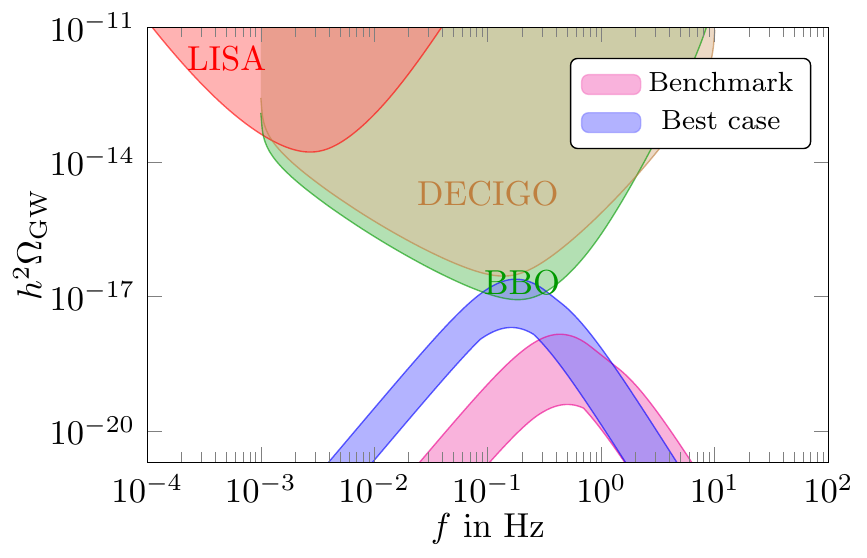}
	\hfill
	\includegraphics[width=.329\linewidth]{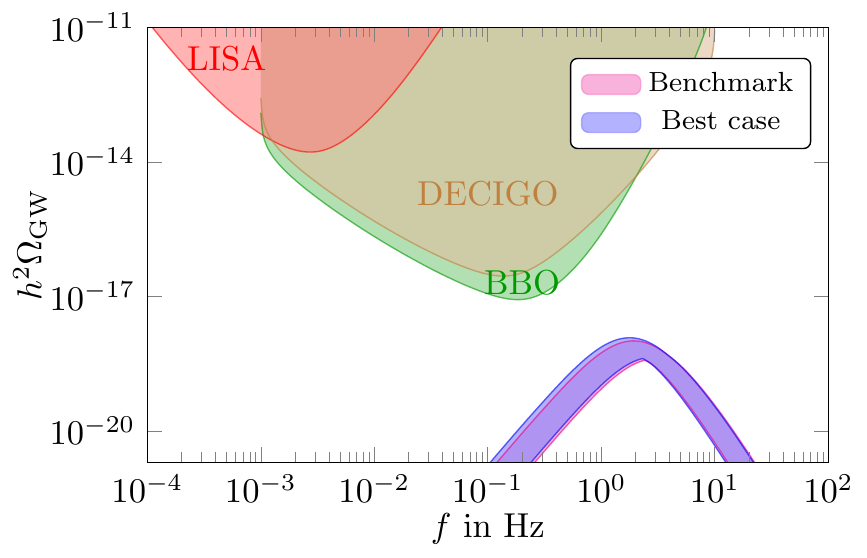}
	\hfill
	\includegraphics[width=.329\linewidth]{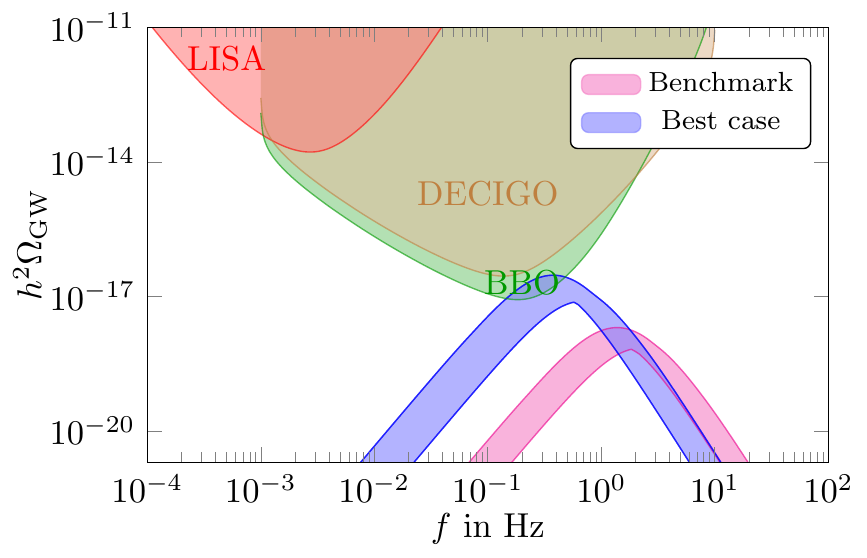}
	\caption{GW spectra of the chiral/confinement phase transition of SU(3) with $N_f= 3$ fundamental quarks (left), $N_f=1$ adjoint quarks (middle), and $N_f = 1$ two-index symmetric quarks (right) for a benchmark scenario and the best-case scenario in the parameter space. Figures taken from~\cite{Reichert:2021cvs}.}
	\label{fig:GW3}
\end{figure}

We now include quarks to SU(3) with the PNJL model, specifically, we include $N_f= 3$ fundamental, $N_f=1$ adjoint, and $N_f = 1$ two-index symmetric quarks. The results are displayed in \cref{fig:GW3} for the benchmark case and the best-case scenario in the parameter space, see \cref{sec:PNJL}. Note that with fundamental quarks, the centre symmetry is broken and it is the first-order chiral phase transition that generates the GWs. With adjoint and two-index symmetric quarks, the centre symmetry is preserved\footnote{Two-index symmetric quarks break the centre symmetry softly and it is thus almost preserved.} and it is the first-order confinement phase transition that generates the GWs. Compared to the pure gluon case, adjoint quarks suppress the GW signal while two-index symmetric quarks enhance it. The chiral phase transition with fundamental quarks is stronger than the pure gluon case, of similar strength as the two-index symmetric case. 

\section{Conclusions and Outlook}
In this contribution, we discussed the prospects of measuring a stochastic GW signal from a first-order chiral or confinement phase transition from a dark sector. We investigated the pure glue SU($N_c$) confinement phase transition for arbitrary $N_c$ as well as at the SU(3) phase transition with $N_f=3$ fundamental quarks (chiral phase transition), with $N_f = 1$ adjoint quarks and $N_f =1$ two-index symmetric quarks (both confinement phase transition). We observe that all GW signals are strongly suppressed due to the large parameter $\beta = \mathcal{O}(10^4)$. Observationally, this implies that DECIGO or BBO will be able to test these dark sectors for confinement temperatures around $T_c \sim \mathcal{O}(10)$\,GeV. Beyond the results presented here, there is a large unexplored dark confined landscape and it will be exciting to see if this trend continues for other strongly coupled dark sectors.

{\bf Acknowledgments} 
We thank W.C.~Huang, R.~Pasechnik, F.~Sannino, and C.~Zhang for joint work on the topics discussed above. MR acknowledges support by the Science and Technology Research Council (STFC) under the Consolidated Grant ST/T00102X/1.

\bibliography{refs}

\end{document}